\documentstyle[12pt,epsfig,rotating]{article}
\begin{document}
\baselineskip=18pt

\hspace{3.7in}Version 2.41, Sept 5, 2002
 

\centerline{\bf Consistently computing the $K\rightarrow\pi$ long
distance weak transition}
 
\vspace{0.15in}

\renewcommand{\thefootnote}{\fnsymbol{footnote}}
 
\centerline{J. Lowe}

\centerline{\it Physics Department, University of New Mexico, Albuquerque,
NM 87131}
 
\centerline{M.D. Scadron}

\centerline{\it Physics Department, University of Arizona, Tucson, 
AZ 85721}
 
\renewcommand{\thefootnote}{\arabic{footnote}}
 
\vspace{0.5in}
 
\begin{abstract}
\noindent First we extract the long-distance (LD)
weak matrix element $\langle\pi\mid H_W\mid K\rangle$ from 
$K\rightarrow 2\pi$, $K\rightarrow 3\pi$, $K_L\rightarrow 2\gamma$,
$K^+\rightarrow\pi^+e^+e^-$, $K^+\rightarrow\pi^+\mu^+\mu^-$ 
and $K_L\rightarrow\mu^+\mu^-$ data and 
give compatible theoretical estimates. We also link this LD scale to 
the single-quark-line (SQL) transition scale $\beta_W$ and then test 
the latter SQL scale against the decuplet weak decay amplitude
ratio $\langle\pi^-\Xi^0\mid H_W\mid\Omega^-\rangle/
\langle\Xi^-\mid H_W\mid\Omega^-\rangle$. Finally, we study LD
$K_S\rightarrow 2\gamma$ decay. All of these experimental and
theoretical values for $\langle\pi\mid H_W\mid K\rangle$ are in
good agreement. We deduce an average value from
eleven experimental determinations of 
$\langle\pi\mid H_W\mid K\rangle=(3.59\pm 0.05)\times 
10^{-8}~{\rm GeV}^2$, while the theoretical SQL values average to
$(3.577\pm 0.004)\times 10^{-8}~{\rm GeV}^2$.
 
\vspace{0.15in}
 
\end{abstract}
  
\vspace{0.35in} 
 

\centerline{\bf I. INTRODUCTION}
 
\vspace{0.05in}
 
The weak matrix element, $\langle\pi\mid H_W\mid K\rangle$, appears
in expressions for many kaon decay rates. Here, we deduce values for 
it from several experimental and theoretical sources. The work is an
up-date to, and extension of, an earlier analysis \cite{ks}. Our 
goal is to demonstrate that there is a good deal of consistent
information, both experimental and theoretical, on this matrix
element, and that a reliable numerical value is available.
In Sect. II, we
extract four values of  $\langle\pi\mid H_W\mid K\rangle$ from data
and we show that these are in good agreement with each other and
with two theoretical estimates. In Sects. III and IV, we study the
SQL scale, $\beta_W$, demonstrating consistency
with values from first and second-order treatments and yielding two
further values for the $K\rightarrow\pi$ weak matrix element.
In Sect. V, we derive yet another value for
$\langle\pi\mid H_W\mid K\rangle$
from the weak $K_L\rightarrow\mu^+\mu^-$ rate together with the
electromagnetic $\pi e\bar{e}$ and $\eta\mu\bar{\mu}$
rates, and in Sect. VI, we demonstrate consistency of our
numerical result with the
parity-violating $K_S\rightarrow\gamma\gamma$ decay. Sect. VII
summarises the results.
 
Throughout, the work uses the current-algebra (CA) partially
conserved axial current (PCAC)-consistency scheme, as was studied
in Ref. \cite{ks}. Such a low-energy CA-PCAC chiral approach
generates long-distance (LD) weak amplitudes for an $H_W$ built
up from $V-A$ weak currents satisfying the charge algebra
$[Q_5+Q,H_W]=0$.
  
Unless otherwise stated, all data in the following are taken from
the Particle Data Group (PDG) tabulation \cite{pdg}.
 
\vspace{1.0cm}
 
{\bf II. EXPERIMENTAL and THEORETICAL TESTS of the $K\rightarrow\pi$
WEAK SCALE}
 
\vspace{0.3cm}
 
\noindent {\bf II.1} From $K_S\rightarrow\pi\pi$ decays
 
\vspace{0.05in}
 
\noindent CA-PCAC consistency requires amplitude magnitudes, 
for $f_{\pi}\approx 93~{\rm MeV}$ \cite{ks}:
 
\begin{eqnarray}
\mid{\cal M}^{+-}_{K_S\pi\pi}\mid &=&\frac{1}{f_{\pi}}
\mid\langle\pi^+\mid H_W\mid K^+\rangle\mid (1-m^2_{\pi}/m^2_K),
\nonumber \\
\mid{\cal M}^{00}_{K_S\pi\pi}\mid &=&\frac{1}{f_{\pi}}
\mid\langle\pi^0\mid H_W\mid K_L\rangle\mid (1-m^2_{\pi}/m^2_K),
\nonumber \\
\mid{\cal M}^{+0}_{K^+\pi^+\pi^0}\mid &=&\frac{1}{2f_{\pi}}
\mid\langle\pi^+\mid H_W\mid K^+\rangle +
\langle\pi^0\mid H_W\mid K_L\rangle\mid
(1-m^2_{\pi}/m^2_K),
\label{eq.1}
\end{eqnarray}
 
\noindent where the factor $(1-m^2_{\pi}/m^2_K)$ must be used 
in Eqs. \ref{eq.1} \cite{cab}. Note that the isospin invariance 
$\langle\pi^+\mid H_W\mid K^+\rangle =-
\langle\pi^0\mid H_W\mid K_L\rangle$ manifests the $\Delta I=1/2$
rule in Eq. \ref{eq.1} with 
$\mid {\cal M}^{+-}\mid=\mid {\cal M}^{00}\mid$
and $\mid {\cal M}^{+0}\mid\rightarrow 0$, then compatible with data
and $\mid {\cal M}^{+0} \mid /\mid {\cal M}^{+-}\mid \approx 0.047$.
However, one must also account for the final-state [FS] interactions
with the observed $\pi\pi$ phase shifts, $\delta^{I_{\pi\pi}}_{\l}$,
evaluated at the kaon mass,
$\delta^0_0 -\delta^2_0 \approx 45^o$ \cite{pipi}. To do so,
the $K_S\rightarrow\pi\pi$ amplitudes of Eqs. \ref{eq.1} are written 
as
 
\begin{eqnarray}
{\cal M}^{+-}_{K_S\pi\pi}=a_{1/2}+\frac{2}{3}a_{3/2}
\nonumber \\
{\cal M}^{00}_{K_S\pi\pi}=a_{1/2}-\frac{4}{3}a_{3/2},
\label{eq.2}
\end{eqnarray}
 
\noindent where the subscripts on the real amplitudes $a_{1/2},$
$a_{3/2}$ refer to the I-spin of $H_W$. The amplitudes including
FS interactions are then \cite{ks}
 
\begin{eqnarray}
\mid{\cal M}^{+-}_{K_S\pi\pi}\mid_{FS}=\left[a^2_{1/2}+
\frac{4}{9}a^2_{3/2}+\frac{4}{3}a_{1/2}a_{3/2}{\rm cos}
(\delta^0_0-\delta^2_0)\right]^{1/2}
\nonumber \\
\mid{\cal M}^{00}_{K_S\pi\pi}\mid_{FS}=\left[a^2_{1/2}+
\frac{16}{9}a^2_{3/2}-\frac{8}{3}a_{1/2}a_{3/2}{\rm cos}
(\delta^0_0-\delta^2_0)\right]^{1/2}.
\label{eq.3}
\end{eqnarray}
 
The experimental rates, from Ref. \cite{pdg}, give amplitudes
 
\begin{eqnarray}
\mid{\cal M}^{+-}_{K_S\pi\pi}\mid_{\rm PDG}&=&m_{K_S}
\left[8\pi\Gamma^{K_S}_{+-}
/q\right]^{\frac{1}{2}}=(39.1\pm 0.2)\times 10^{-8}~{\rm GeV},
\nonumber \\
\mid{\cal M}^{00}_{K_S\pi\pi}\mid_{\rm PDG}&=&m_{K_S}
\left[16\pi\Gamma^{K_S}_{00}
/q\right]^{\frac{1}{2}}=(37.1\pm 0.1)\times 10^{-8}~{\rm GeV},
\nonumber \\
\mid{\cal M}^{+0}_{K^+\pi\pi}\mid_{\rm PDG}&=&m_{K^+}
\left[8\pi\Gamma^{K^+}_{+0}
/q\right]^{\frac{1}{2}}=(1.832\pm 0.006)\times 10^{-8}~{\rm GeV},
\label{eq.4}
\end{eqnarray}
 
\noindent where $q$ is the magnitude of the relevant decay momentum.
These can be substituted in Eqs. \ref{eq.3} yielding the I-spin 
amplitudes

\begin{eqnarray}
a_{1/2}\approx 38.42\times 10^{-8}~{\rm GeV},
\nonumber \\
a_{3/2}\approx 1.43\times 10^{-8}~{\rm GeV}.
\label{eq.5}
\end{eqnarray}

\begin{figure}[h]
\hspace{3.0cm}
\epsfig{file=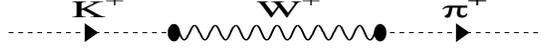,height=6.0cm,width=12.0cm}
\vspace{-1.0cm}
\caption{{\it W pole graph for the transition}
$\langle\pi^+\mid H_W\mid K^+\rangle$.\label{fig.1}} 
\end{figure}

\noindent These can in turn be used in 
the CA-PCAC consistency conditions, 
Eqs. \ref{eq.1}, \ref{eq.2}, to extract
the reduced matrix element $\langle\pi^+\mid H_W\mid K^+\rangle$
(or equivalently $-\langle\pi^0\mid H_W\mid K_L\rangle$).
For $\langle\pi^+\mid H_W\mid K^+\rangle$, it is necessary to
{\it subtract} off the contribution from 
the $W^+$ pole graph of Fig. \ref{fig.1} \cite{wpole}, namely
 
\vspace{-1.0cm}
\vspace{1.0cm}

\begin{eqnarray}
\mid\langle\pi^+\mid H_W\mid K^+\rangle\mid_{\rm W pole}=
(G_F/\sqrt{2})s_1c_1f_{\pi}f_Km^2_K
=(0.460\pm 0.006)\times 10^{-8}~{\rm GeV}^2
\label{eq.6}
\end{eqnarray}
 
\noindent for $G_F=11.6639\times 10^{-6}~{\rm GeV}^{-2}$,  
$f_K\approx 1.22f_{\pi}$, $f_{\pi}\approx 93~{\rm MeV}$ and 
$s_1c_1= 0.217\pm 0.003$ \cite{pdg}. This leads to the result
\cite{pol}
  
\begin{eqnarray}
\mid\langle\pi^+\mid H_W\mid K^+\rangle\mid
&=&((3.980\pm 0.020)-(0.460\pm 0.006))\times 10^{-8}~{\rm GeV}^2
\nonumber \\
&=&(3.520\pm 0.021)\times 10^{-8}~{\rm GeV}^2,
\label{eq.7}
\end{eqnarray}
 
\noindent where the first 3.980 number in Eq. \ref{eq.7} stems from 
Eqs. \ref{eq.1}-\ref{eq.5}, $\mid{\cal M}^{+-}_{K_S\pi\pi}\mid\approx
39.373\times 10^{-8}~{\rm GeV}$. For the matrix element 
$\langle\pi^0\mid H_W\mid K_L\rangle$, for which there is no 
W pole term, Eqs. \ref{eq.1}-\ref{eq.3} give directly
 
\begin{eqnarray}
\mid\langle\pi^0\mid H_W\mid K_L\rangle\mid
=(3.666\pm 0.010)\times 10^{-8}~{\rm GeV}^2.
\label{eq.8}
\end{eqnarray}

\noindent It is reassuring that these values in Eqs. \ref{eq.7} and 
\ref{eq.8} are quite close.
Again assuming isospin invariance, we average Eqs. \ref{eq.7} 
and \ref{eq.8} to find the mean reduced matrix element
 
\begin{eqnarray}
\mid\langle\pi^+\mid H_W\mid K^+\rangle\mid_{\rm avg}~=~
\mid\langle\pi^0\mid H_W\mid K_L\rangle\mid_{\rm avg}~=~
(3.637\pm 0.009)\times 10^{-8}~{\rm GeV}^2.
\label{eq.9}
\end{eqnarray}
 
\vspace{1.0cm}
 
\noindent {\bf II.2} From $K\rightarrow 3\pi$ decays
 
Reducing in two pions consistently, the four measured $K_{3\pi}$ weak
decay amplitudes \cite{pdg} 
 
\noindent ($\mid{\cal M}^{+}_{++-}\mid=(1.93\pm 0.01)\times 10^{-6}$,
$\mid{\cal M}^{+}_{+00}\mid=(0.96\pm 0.01)\times 10^{-6}$,
$\mid{\cal M}^{L}_{+0-}\mid=(0.91\pm 0.01)\times 10^{-6}$,
$\mid{\cal M}^{L}_{000}\mid=(2.60\pm 0.02)\times 10^{-6}$) predict the
CA-PCAC reduced matrix elements \cite{ks}:
 
\begin{eqnarray}
\mid\langle\pi^+\mid H_W\mid K^+\rangle\mid&=&2f_{\pi}^2
(1-m^2_{\pi}/m^2_K)^{-1}
\mid\langle\pi^+\pi^+\pi^-\mid H_W\mid K^+\rangle\mid
=(3.63\pm 0.02)\times 10^{-8}~{\rm GeV^2},
\nonumber \\
\mid\langle\pi^+\mid H_W\mid K^+\rangle\mid&=&4f_{\pi}^2
(1-m^2_{\pi}/m^2_K)^{-1}
\mid\langle\pi^+\pi^0\pi^0\mid H_W\mid K^+\rangle\mid
=(3.59\pm 0.04)\times 10^{-8}~{\rm GeV^2},
\nonumber \\
\mid\langle\pi^0\mid H_W\mid K_L\rangle\mid&=&4f_{\pi}^2
(1-m^2_{\pi}/m^2_K)^{-1}
\mid\langle\pi^+\pi^0\pi^-\mid H_W\mid K_L\rangle\mid
=(3.42\pm 0.04)\times 10^{-8}~{\rm GeV^2},
\nonumber \\
\mid\langle\pi^0\mid H_W\mid K_L\rangle\mid&=&\frac{4}{3}f_{\pi}^2
(1-m^2_{\pi}/m^2_K)^{-1}
\mid\langle 3\pi^0\mid H_W\mid K_L\rangle\mid
=(3.24\pm 0.02)\times 10^{-8}~{\rm GeV^2}.
\label{eq.10}
\end{eqnarray}
 
\noindent Once more assuming isospin invariance, the four reduced 
$K\rightarrow\pi$ transitions for $K_{3\pi}$ decays in Eq. \ref{eq.10}  
average to 
 
\begin{eqnarray}
\mid\langle\pi^+\mid H_W\mid K^+\rangle\mid^{K_{3\pi}}_{\rm PCAC}=
\mid\langle\pi^0\mid H_W\mid K_L\rangle\mid^{K_{3\pi}}_{\rm PCAC}
=(3.449\pm 0.015)\times 10^{-8}~{\rm GeV^2}.
\label{eq.11}
\end{eqnarray}
 
To show that the final-state-corrected $K_{2\pi}$ result, Eq. \ref{eq.9} 
and the nearby PCAC-averaged $K_{3\pi}$ result, Eq. \ref{eq.11}, (where FS
interactions are known \cite{pipi} to be minimal) do indeed refer
to the same $K\rightarrow\pi$ weak transition, we must verify that
the PCAC corrections in the two cases are in fact minimal, given
the observed rate ratio \cite{pdg}:
 
\begin{eqnarray}
\frac{\Gamma(K^+\rightarrow\pi^+\pi^+\pi^-)}
{\Gamma(K_S\rightarrow\pi^+\pi^-)}\bigg|_{\rm PDG}=
\frac{(\hbar/\tau_{K^+})(5.59\pm 0.08)\%}
     {(\hbar/\tau_{K_S})(68.61\pm 0.28)\%}=
(5.88\pm 0.09)\times 10^{-4}.
\label{eq.12}
\end{eqnarray}
 
\noindent This ratio, Eq. \ref{eq.12}, must then be compared to the analog
PCAC rate ratio, where the PCAC-consistency amplitude ratio \cite{ks}  
$\mid A^+_{++}\mid_{PCAC}=\mid\langle\pi^+\pi^-\mid H_W\mid K_S\rangle
\mid/2f_{\pi}$ generates the rate ratio
 
\begin{eqnarray}
\frac{\Gamma(K^+\rightarrow\pi^+\pi^+\pi^-)}
{\Gamma(K_S\rightarrow\pi^+\pi^-)}\bigg|_{\rm PCAC}=
\frac{0.798\times 10^{-6}~{\rm GeV}}{0.206~{\rm GeV}}
\left(\frac{1}{2f_{\pi}}\right)^28\pi m^2_{K_S}
\left(1-\frac{m^2_{\pi}}{m^2_K}\right)^2
\approx 5.94\times 10^{-4}.
\label{eq.13}
\end{eqnarray}
 
\noindent Note that the measured rate ratio Eq. \ref{eq.12} is close to
the PCAC-consistency ratio Eq. \ref{eq.13}. Here, the 3-body phase-space 
factor 
$0.798\times 10^{-6}$ GeV was computed in Ref. \cite{ks}. Likewise
the rate ratio for $K_L\rightarrow 3\pi^0/K_S\rightarrow 2\pi^0$ is
for the PDG $(1.16\pm 0.02)\times 10^{-3}$ and $1.32\times 10^{-3}$
for PCAC consistency. The amplitude ratios are even
closer. Thus it should not be surprising that the scales of 3.64 in
Eq. \ref{eq.9} and 3.45 in Eq. \ref{eq.11} are indeed close.
 
\vspace{0.6cm}
  
\noindent {\bf II.3} From $K_L\rightarrow 2\gamma$ decays
 
\begin{figure}[!h!]
\hspace{2.0cm}
\epsfig{file=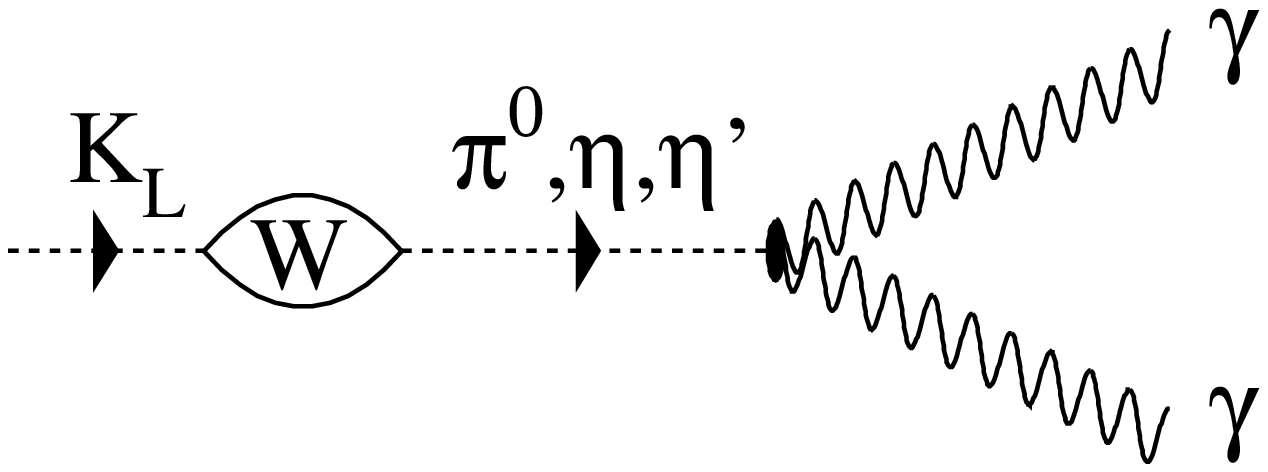,height=7.0cm,width=12.0cm}
\vspace{-1.0cm}
\caption{{\it Meson} $\pi^0,\eta,\eta'$ {\it pole graphs for the}
$K_L\rightarrow 2\gamma$ {\it decay.}\label{fig.2}} 
\end{figure}
\vspace{1.0cm}
  
Low-mass pole diagrams also generate LD weak amplitudes.
The $\pi^0$ pole gives the $K_L\rightarrow 2\gamma$ amplitude
magnitude \cite{pdg} 
 
\begin{eqnarray}
\mid\langle 2\gamma\mid H_W\mid K_L\rangle\mid=
\frac{\mid\langle\pi^0\mid H_W\mid K_L\rangle\mid}
{m^2_{K_L}-m^2_{\pi^0}}
F_{\pi^0\gamma\gamma}=(3.49\pm 0.05)\times 10^{-9}~{\rm GeV}^{-1},
\label{eq.14}
\end{eqnarray}
 
\noindent where the $\pi^0\gamma\gamma$ amplitude (either from data
or theory) is $\alpha/\pi f_{\pi}=0.025~{\rm GeV}^{-1}$ and we have 
neglected the Levi-Civita factor on both sides of Eq. \ref{eq.14}. 
This then leads immediately to 
 
\begin{eqnarray}
\mid\langle\pi^0\mid H_W\mid K_L\rangle\mid_{\pi^0~{\rm pole}}~=~
(3.20\pm 0.04)\times 10^{-8}~{\rm GeV}^2.
\label{eq.15}
\end{eqnarray}
 
\noindent The $\eta$ and $\eta'$ pole graphs in Fig. \ref{fig.2} have 
opposite sign. However, they are not quite equal in magnitude and 
a detailed calculation \cite{pol} shows that the 
$K\rightarrow\pi$ transition in Eq. \ref{eq.15} is then {\it effectively 
enhanced} by 11.1\% to
  
\begin{eqnarray}
\mid\langle\pi^0\mid H_W\mid K_L\rangle
\mid_{\pi^0,\eta,\eta'{\rm poles}}~
=(3.56\pm 0.04)\times 10^{-8}~{\rm GeV}^2.
\label{eq.16}
\end{eqnarray}
 
\vspace{1.0cm}
 
\noindent {\bf II.4} From the $K^+\rightarrow \pi^+e^+e^-$ and
$K^+\rightarrow \pi^+\mu^+\mu^-$ rare decays
 
The recent Brookhaven $K^+\rightarrow \pi^+e^+e^-$ experiment E865 finds
the amplitude \cite{865,bels} at $q^2=0$
 
\begin{eqnarray}
\mid A(0)\mid_{K\pi e\bar{e}}=
(4.00\pm 0.18)\times 10^{-9}~{\rm GeV}^{-2}.
\label{eq.17}
\end{eqnarray}
  
\vspace{-1.0cm}
\begin{figure}[h]
\hspace{1.0cm}
\epsfig{file=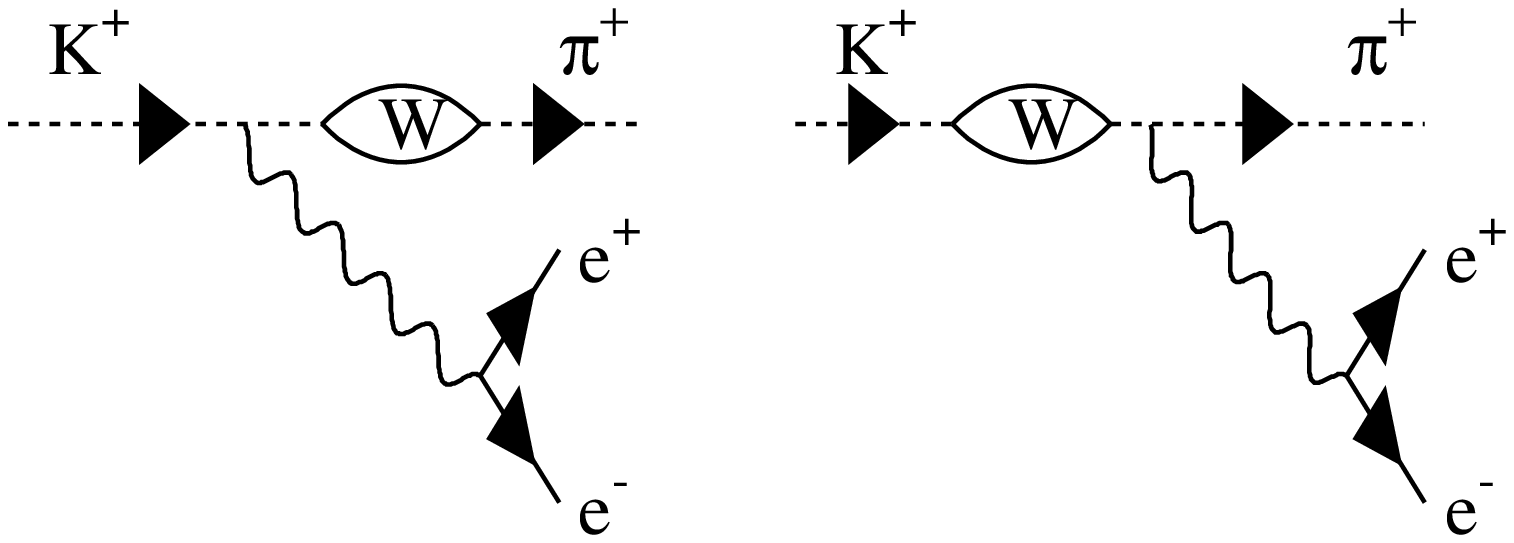,height=7.6cm,width=12.0cm}
\vspace{-1.5cm}
\caption{{\it Virtual bremsstrahlung graphs for}
$K^+\rightarrow \pi^+e^+e^-$ {\it decay.}\label{fig.3}} 
\end{figure}
\vspace{1.5cm}
  
\noindent Recently, Burkhardt {\it et al.} \cite{bels} showed that
both the decay rate and the $q^2$ dependence can be well
understood in a straightforward model in which the only terms that 
survive at $q^2=0$ are the LD virtual bremsstrahlung graphs of
Fig. \ref{fig.3}, which scale with the weak matrix element 
$\langle\pi^+\mid H_W\mid K^+\rangle$ that we study here.
These virtual bremsstrahlung graphs predict the 
long-distance (LD) weak amplitude at $q^2=0$ \cite{bels}:
 
\begin{eqnarray}
\mid A_{LD}(0)\mid=e^2\bigg|\frac{\langle\pi^+\mid H_W\mid K^+\rangle}
{m^2_{K^+}-m^2_{\pi^+}}\bigg|~\bigg|\frac{dF_{\pi^+}}{dq^2} -
\frac{dF_{K^+}}{dq^2}\bigg|,
\label{eq.18}
\end{eqnarray}
 
\noindent where vector meson dominance (VMD) $\rho$, $\omega$ and 
$\phi$ poles require e.g. the $\rho$ form factor 
$F_{\pi^+}(q^2)=\left(1-q^2/m^2_{\rho}\right)^{-1}$ and
 
\begin{eqnarray}
\frac{dF_{\pi^+}}{dq^2}\bigg|_{q^2=0}=1/m^2_{\rho}=1.69~{\rm GeV}^{-2}.
\label{eq.19}
\end{eqnarray}
 
\noindent Likewise, the $\rho$, $\omega$ and $\phi$ VMD poles for
the $F_{K^+}(q^2)$ form factor predict \cite{bels}
 
\begin{eqnarray}
\frac{dF_{K^+}}{dq^2}\bigg|_{q^2=0}=\frac{n_{\rho}}{m^2_{\rho^0}}+
\frac{n_{\omega}}{m^2_{\omega}}+\frac{n_{\phi}}{m^2_{\phi^0}}=
1.42~{\rm GeV}^{-2}
\label{eq.20}
\end{eqnarray}
 
\noindent as measured from $\rho,\omega,\phi\rightarrow e^+e^-$
electromagnetic decays. Here, $6n_{\rho,\omega,\phi}=1,3,2$ from
VMD \cite{bels,ayala} fixing the normalisation 
$n_{\rho}+n_{\omega}+n_{\phi}=1$ as required. Finally, substituting
Eqs. \ref{eq.19}, \ref{eq.20} into Eq. \ref{eq.18} and using the E865 
$K^+\rightarrow\pi^+e^+e^-$ amplitude in Eq. \ref{eq.17}, one extracts
the LD transition:
 
\begin{eqnarray}
\mid\langle\pi^+\mid H_W\mid K^+\rangle\mid&=&e^{-2}
\mid A(0)\mid_{K^+\rightarrow\pi^+e^+e^-}
\left(m^2_{K^+}-m^2_{\pi^+}\right)
\bigg|\frac{dF_{\pi^+}}{dq^2}-\frac{dF_{K^+}}{dq^2}\bigg|^{-1}_{q^2=0}
\nonumber \\
&=&(3.62\pm 0.16)\times 10^{-8}~{\rm GeV}^2.
\label{eq.21}
\end{eqnarray}
 
\noindent Possible short-distance (SD) corrections to Eq. \ref{eq.21} 
have been
shown to be less than 10\% \cite{ddg}, partly because the top quark 
corrections are small due to the large top-quark mass
($m_t\sim 175 ~{\rm MeV}$) and also because QCD corrections 
substantially reduce the SD contribution. This lack of knowledge
of the SD contribution is the dominant source of uncertainty
in the weak matrix element, so we take 
the value to be

\begin{eqnarray}
\mid\langle\pi^+\mid H_W\mid K^+\rangle\mid=
(3.62\pm 0.36)\times 10^{-8}~{\rm GeV}^2.
\label{eq.22}
\end{eqnarray}
  
\noindent Eq. \ref{eq.22} is in close agreement with Eqs. \ref{eq.9},
\ref{eq.11}, \ref{eq.16} above. 
 
Brookhaven experiment E865 has also measured \cite{865pmm} the decay 
$K^+\rightarrow\pi^+\mu^+\mu^-$. Because of the limited 
statistics of this measurement, compared to the 
$K^+\rightarrow\pi^+ e^+ e^-$ channel, an
extrapolation quadratic in $q^2$ gives an unduly large
error in the amplitude at $q^2=0$. Therefore, we invoke
lepton universality which implies that the shape of the 
$q^2$ dependence of $A(q^2)$ is the same for these two channels.
Thus we fit the amplitude, as a function of $q^2$, for 
$K^+\rightarrow\pi^+\mu^+\mu^-$ using the coefficients
of $q^2$ and $q^4$ determined from the fit \cite{865} to the 
$K^+\rightarrow\pi^+ e^+ e^-$ data. Again, we allow a 
contribution of 10\% to the error in $A(0)$ due to the 
uncertainty in the SD contribution. The result is
 
\begin{eqnarray}
\mid A(0)\mid_{K\pi\mu\bar{\mu}}=(4.45\pm 0.67)\times 
10^{-9}~{\rm GeV}^{-2}
\label{eq.23}
\end{eqnarray}
 
\noindent from which, using Eqs. \ref{eq.18} and \ref{eq.21},

\begin{eqnarray}
\mid\langle\pi^+\mid H_W\mid K^+\rangle\mid=
(4.0\pm 0.6)\times 10^{-8}~{\rm GeV}^2.
\label{eq.24}
\end{eqnarray}

\vspace{1.0cm}
 
\noindent {\bf II.5} Meson weak self-energy graphs
 
\begin{figure}[h]
\hspace{3.0cm}
\epsfig{file=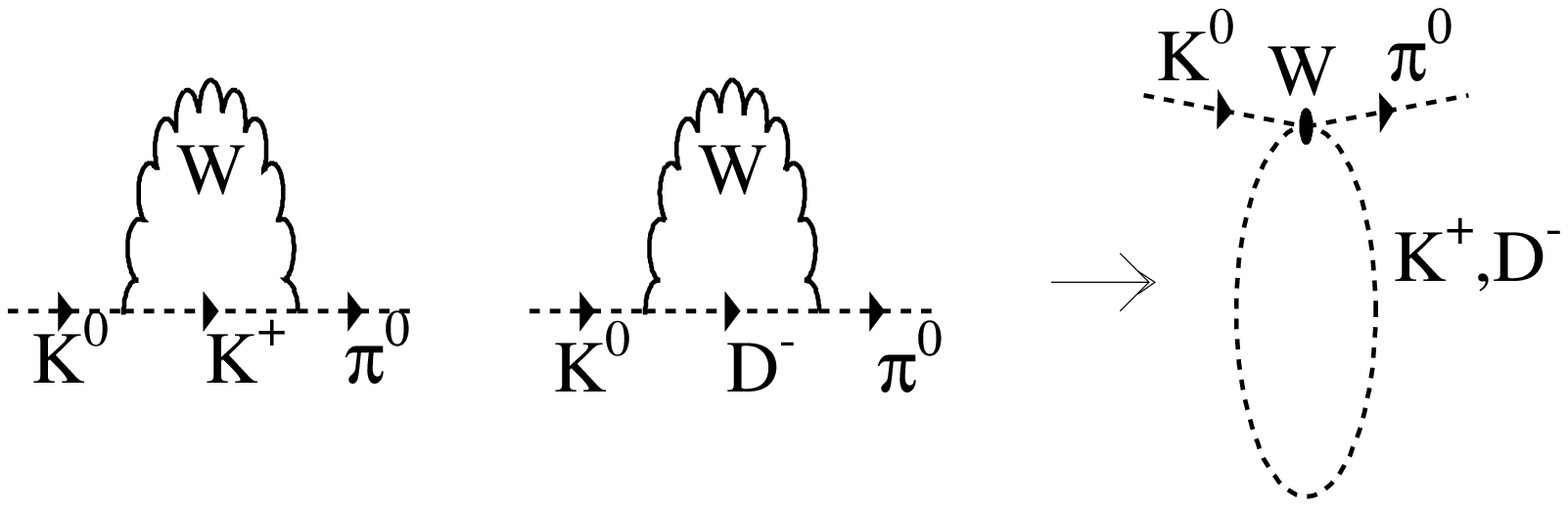,height=5.5cm,width=11.0cm}
\caption{{\it Meson W-mediated self-energy type graphs for the}
$\langle\pi^0\mid H_W\mid K^0\rangle$ {\it transition.}} 
\label{fig.4}
\end{figure}

We now compare the five experimental $\langle\pi\mid H_W\mid 
K\rangle$ LD weak scales in Eqs. \ref{eq.9}, \ref{eq.11}, \ref{eq.16},
\ref{eq.22}, \ref{eq.24} (extracted from $K_{2\pi}$, 
$K_{3\pi}$, $K_{L2\gamma}$, $K^+_{\pi^+e^+e^-}$, 
$K^+_{\pi^+\mu^+\mu^-}$ weak decays, averaging to 
$3.59\times 10^{-8}~{\rm GeV}^2$) 
with the theoretical prediction of
the ``eye diagram" model-independent meson loop graphs of Fig. \ref{fig.4}.

\noindent This amplitude, which has UV cutoff 
$\Lambda\approx 1.87~{\rm GeV}$
near the observed D mass \cite{ks2}, is found {\it via} a Wick
rotation to $q^2=-p^2$, $d^4p=i\pi^2q^2dq^2$
with $q^2=x$ and $s_1c_1=0.217\pm 0.003$:
 
\begin{eqnarray}
\mid\langle\pi^0\mid H_W\mid K_L\rangle\mid&=&\frac{G_F}{\sqrt{2}}
s_1c_1\left(m^2_D-m^2_K\right)\int^{\Lambda}\frac{d^4p}{(2\pi)^4}
\frac{p^2}{(p^2-m^2_D)(p^2-m^2_K)}
\nonumber \\
&=&\frac{G_FV_{ud}V_{us}\left(m^2_D-m^2_K\right)}{\sqrt{2}\times 16\pi^2}
\int^{\Lambda^2}_0\frac{x^2~dx}{(x+m_D^2)(x+m_K^2)}
\nonumber \\
&=& (3.47\pm 0.05)\times 10^{-8}~{\rm GeV}^2,
\label{eq.25}
\end{eqnarray}
 
\noindent where the error is dominated by the uncertainty in $s_1c_1$.
This result is in excellent agreeent with the weak LD scales in Eqs.
\ref{eq.9}, \ref{eq.11}, \ref{eq.16}, \ref{eq.22}, \ref{eq.24} above.
 
An alternative W-mediated approach which builds in the
$\Delta I=1/2$ structure follows from the quark-model 
single-quark-line (SQL) 
s$\rightarrow$d weak transition. With u and c quark intermediate 
states, the GIM $K_L\rightarrow\pi^0$ LD transition 
is \cite{delsca}
 
\begin{eqnarray}
\mid\langle\pi^0\mid H_W\mid K_L\rangle\mid=\frac{G_F}{\sqrt{2}}
\frac{s_1c_1}{4\pi^2}(m^2_c-m^2_u)m^2_K\frac{f_K}{f_{\pi}}
\approx 2.92\times 10^{-8}~{\rm GeV}^2
\label{eq.26}
\end{eqnarray}
 
\noindent for $G_F=11.6639\times 10^{-6}~{\rm GeV}^{-2}$,
$s_1c_1\approx 0.217$, $m_c\approx 1.5$ GeV, $m_u\approx 0.34$ GeV
(for constituent quarks) and $f_K/f_{\pi}\approx 1.22$. Then
the predicted Eq. \ref{eq.26} is 82\% of Eqs. (9,11,16,22,24) above. 
When the heavier top-quark intermediate state in included using a
``heavy-quark approximation" \cite{delliu}, Eq. \ref{eq.26} becomes 
closer to the $K_L\rightarrow\pi^0$ LD transition found throughout
sect. II.

\vspace{1.0cm}
 
\noindent {\bf III. FIRST and SECOND ORDER TESTS of the
SQL TRANSITION}
 
\vspace{0.25in}

\noindent {\bf III.1} Quark-model $s\rightarrow d$ first-order 
weak transition
 
\begin{figure}[h]
\hspace{2.0cm}
\epsfig{file=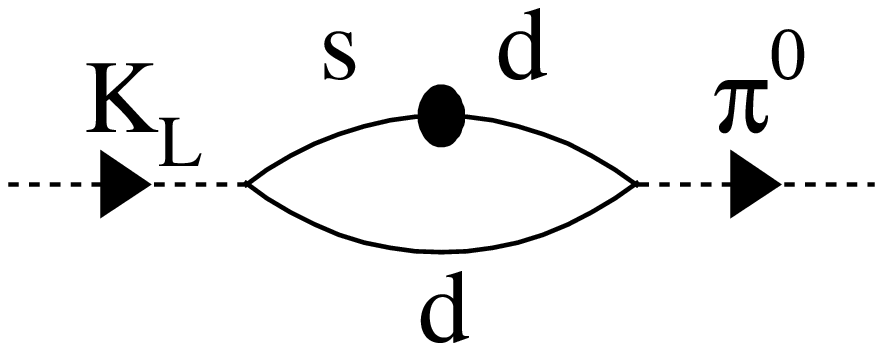,height=9.0cm,width=12.0cm}
\vspace{-4.0cm}
\caption{{\it Quark-model} s$\rightarrow$d {\it first-order weak transition
characterising }$\langle\pi^0\mid H_W\mid K_L\rangle$.\label{fig.5}} 
\end{figure}

Alternatively, we study the $K_L\rightarrow \pi^0$ weak transition
at the quark level {\it via} the (Nambu-Goldstone) tightly bound 
$\Delta I=\frac{1}{2}$ quark bubble graph of Fig. \ref{fig.5}.
  
\noindent We focus on the $K_S\rightarrow\pi\pi$ decay rate
$\Gamma_S$. Firstly, one computes from Fig. \ref{fig.5} \cite{811}
 
\begin{eqnarray}
\mid\langle\pi^0\mid H_W\mid K_L\rangle\mid =2\beta_W
m^2_{K_L}f_K/f_{\pi},
\label{eq.27}
\end{eqnarray}
 
\noindent where $\beta_W$ is the dimensionless weak SQL scale. 
>From this, the total width of the $K_S$ is \cite{811,fs,ds}
 
\begin{eqnarray}
\Gamma_S\approx \frac{3}{16\pi}\frac{q}{m^2_K}
\mid{\cal M}^{00}_{K_S\pi\pi}\mid^2\approx 
3.61\beta^2_Wm_K.
\label{eq.28}
\end{eqnarray}
 
>From Ref. \cite{pdg}, the experimental value of $\Gamma_S$ is

\begin{eqnarray}
\Gamma_S=(73.67\pm 0.07)\times 10^{-16}~{\rm GeV}.
\label{eq.29}
\end{eqnarray}
 
\noindent Relating Eqs. \ref{eq.28} and \ref{eq.29} gives the weak 
SQL scale as
 
\begin{eqnarray}
\mid\beta_W\mid\approx \left[(73.67\pm 0.07)\times 10^{-16}/3.61\times
0.497672\right]^{1/2}\sim 6.4\times 10^{-8},
\label{eq.30}
\end{eqnarray}
 
\noindent where 0.497672 GeV is the neutral kaon mass. 
Substituting the value of $\beta_W$ from Eq. \ref{eq.30} gives for the
weak matrix element (see the second reference in \cite{bels})
 
\begin{eqnarray}
\mid\langle\pi^0\mid H_W\mid K_L\rangle\mid \sim
3.9\times 10^{-8}~{\rm GeV}^2.
\label{eq.31}
\end{eqnarray} 
 
\noindent This global estimate is compatible with the experimental 
values found in sect. II. We therefore extend the concept of fig. 
\ref{fig.5}
to the second-order weak process, $\Delta m_{LS}=m_{K_L}-m_{K_S}.$
 
\vspace{1.0cm}
 
\noindent {\bf III.2} Quark model s$\rightarrow$d second-order 
weak transition and the $K_L-K_S$ mass difference $\Delta m_{LS}$
 
Since $\Delta m_{LS}$  can be taken as a {\it second-order
weak transition}, one may express the data as \cite{pdg,mrr}
 
\begin{eqnarray}
\Delta m_{LS}=(0.4736\pm 0.0012)\Gamma_S,
{\rm ~or~equivalently}
\label{eq.32}
\end{eqnarray}
 
\begin{eqnarray}
\Delta m_{LS}=(0.7011\pm 0.0016)\times 10^{-14}m_K,
\label{eq.33}
\end{eqnarray}
 
\noindent in order to find a more accurate value for the 
weak SQL scale $\beta_W$ and a further value for 
$\mid\langle\pi^0\mid H_W\mid K_L\rangle\mid$.

\begin{figure}[h]
\hspace{1.5cm}
\epsfig{file=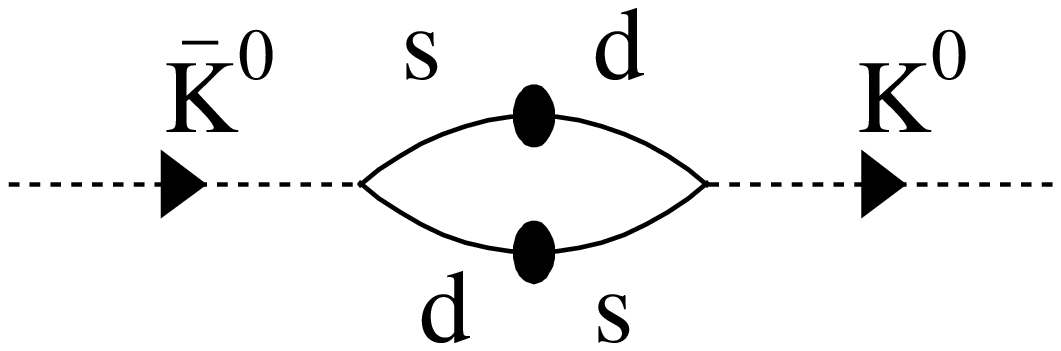,height=9.0cm,width=13.0cm}
\vspace{-4.0cm}
\caption{{\it Quark-model} s$\rightarrow$d {\it second-order 
weak transition
characterising} $\lambda=\langle K^0\mid H_W\mid \bar{K}^0\rangle$.
\label{fig.6}} 
\end{figure}
 
Specifically, the second-order weak quark bubble graph 
of Fig. \ref{fig.6} (not the usual parameter-dependent W-W quark 
box) is the obvious generalisation of the first-order weak quark 
bubble graph of Fig. \ref{fig.5}. In both cases the linear-$\sigma$-model 
inspired (with
PCAC-compatible \cite{gml}) pseudoscalar couplings are used. Then
the (CP-conserving) $K^0-\bar{K}^0$ mixing matrix is diagonalised
as
 
\begin{equation}
\left( \begin{array}{cc}
m^2_{K^0} & \lambda \\
\lambda & m^2_{\bar{K}^0}
\end{array} \right)
\stackrel{\phi}{\rightarrow}
\left( \begin{array}{cc}
m^2_{K_S} & 0 \\
0 & m^2_{K_L}
\end{array} \right),
\label{eq.34}
\end{equation}
 
\noindent with $\phi=45^o$ for states 
$\sqrt{2}\mid K_{L,S}\rangle=\mid K^0\rangle\pm\mid\bar{K}^0\rangle$.
Saturating $\lambda=\langle K^0\mid H_W^{(2)}\mid\bar{K}^0\rangle$
{\it via} unitarity for the overwhelmingly dominant $2\pi$
intermediate state \cite{mrr} predicts 
$\phi={\rm arctan}2\Delta m_{LS}/\Gamma_S\approx\pi/4$, which in turn
gives $\Delta m_{LS}\approx\Gamma_S/2$, close to the measured value
in Eq. \ref{eq.32}.
 
For the above consistent picture (Figs. \ref{fig.5}, \ref{fig.6}), 
ordinary two-level 
quantum mechanics (independent of quantum field theory) then
requires \cite{escs}
 
\begin{eqnarray}
{\rm sin}2\phi=2\lambda(m^2_{K_L}-m^2_{K_S})^{-1}\approx
\lambda/m_K\Delta m_{LS},
\label{eq.35}
\end{eqnarray}
 
\noindent where the LHS of Eq. \ref{eq.35} is unity for $\phi=45^o$ 
(and very nearly unity when CP-violating effects are included)
so that
 
\begin{eqnarray}
\lambda\approx m_K\Delta m_{LS}.
\label{eq.36}
\end{eqnarray}
 
\noindent However, the off-diagonal matrix element $\lambda$ in
Eqs. \ref{eq.35}, \ref{eq.36} also gives from Fig. \ref{fig.6}
 
\begin{eqnarray}
\lambda=\langle K^0\mid H_W^{(2)}\mid \bar{K}^0\rangle
\approx 2\beta^2_Wm^2_K,
\label{eq.37}
\end{eqnarray}
 
\noindent since the only mass scale in Eq. \ref{eq.37} is $m^2_K$,
the diagonalisation is second-order weak (requiring a $\beta^2_W$
factor) and the factor of 2 in Eqs. \ref{eq.27}, \ref{eq.37} stems from 
$(1-\gamma_5)^2=2(1-\gamma_5)$. Standard CA-PCAC considerations
\cite{escs} also recover Eq. \ref{eq.37}.
 
Combining Eqs. \ref{eq.36}, \ref{eq.37} leads to the result
 
\begin{eqnarray}
\Delta m_{LS}\approx 2\beta^2_Wm_K,
\label{eq.38}
\end{eqnarray}
 
\noindent which, together with the $\Delta m_{LS}$ value \cite{pdg}
in Eq. \ref{eq.33}, predicts the weak SQL scale 
 
\begin{eqnarray}
\mid\beta_W\mid= [(0.7011\pm 0.0016)\times 10^{-14}/2]^{1/2}\approx 
(5.921\pm 0.007)\times 10^{-8},
\label{eq.39}
\end{eqnarray}
 
\noindent which gives, with Eq. \ref{eq.27},
 
\begin{eqnarray}
\mid\langle\pi^0\mid H_W\mid K_L\rangle\mid =
(3.578\pm 0.004)\times 10^{-8}~{\rm GeV}^2,
\label{eq.40}
\end{eqnarray} 
 
\noindent again compatible with the experimental values found in
sect. II.

\vspace{1.0cm}

\noindent {\bf IV. PCAC SPIN-3/2 TESTS of the LD-SQL
TRANSITION for} $\langle\Xi\mid H_W\mid\Omega\rangle$
 
\begin{figure}[h]
\hspace{2.8cm}
\epsfig{file=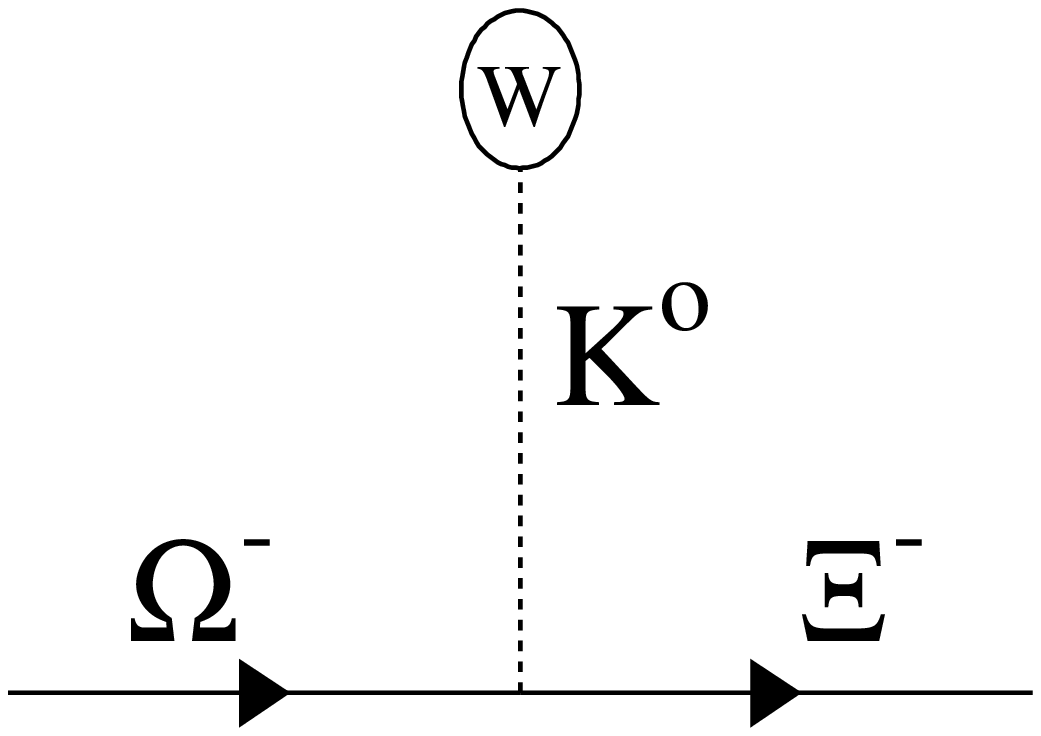,height=7.5cm,width=10.0cm}
\vspace{-2.0cm}
\caption{{\it Kaon tadpole graph characterising }
$\langle\Xi^-\mid H_W\mid \Omega^-\rangle$.\label{fig.7}}
\end{figure}

Another SQL test relates to the decuplet weak decay amplitude
$\Omega^-\rightarrow \Xi^0\pi^-$ relative to the Rarita-Schwinger 
transition $\langle\Xi^-\mid H_W\mid\Omega^-\rangle\approx
h_2\bar{u}(\Xi)p^{\mu}u(\Omega)_{\mu}$, but first we extract this
latter weak scale from the $K^0$ tadpole SQL graph 
of fig. \ref{fig.7}
\cite{uppal} with kaon PCAC strong coupling \cite{scathe}
$\langle K^0\Xi^-\mid\Omega^-\rangle\rightarrow g_2\rightarrow
\sqrt{2}/f_K$:
 
\begin{eqnarray}
\mid\langle\Xi^-\mid H_W\mid\Omega^-\rangle\mid\rightarrow
\mid h_2\mid&=&\mid\langle 0\mid H_W\mid K^0\rangle g_2\mid/m^2_K
\nonumber \\
&\approx &\mid 2\sqrt{2}f_Km^2_K\beta_W\times \sqrt{2}/f_Km^2_K\mid
=4\mid\beta_W\mid.
\label{eq.41}
\end{eqnarray}

\noindent However, kaon PCAC is known to be accurate only to within
25 - 30\%, so instead we employ strong decuplet-octet 
baryon-pseudoscalar meson (DBP) data to extract \cite{scathe}
$g_{DPB}=(m_D+m_B)g_2/2\approx 15.7$ which is a 20\% lower estimate
than used for $g_2$ in Eq. \ref{eq.41}. Thus a more accurate estimate 
for $h_2$ than Eq. \ref{eq.41} is
 
\begin{eqnarray}
\mid h_2\mid=\mid 4\beta_W\mid/1.2\approx 3.3\mid\beta_W\mid .
\label{eq.42}
\end{eqnarray}
 
\noindent We note that Eq. \ref{eq.42} is quite close to the 
standard decuplet-octet baryon SU(3) flavour SQL estimate
 
\begin{eqnarray}
\mid h_2\mid=3\mid\beta_W\mid
\label{eq.43}
\end{eqnarray}
 
\noindent for $\Omega$ constructed from 3 strange quarks. This too
can be verified from magnetic-moment data \cite{pdg}, giving the ratio
 
\begin{eqnarray}
\frac{\mu_{\Omega}}{\mu_{\Lambda}}=
\frac{(2.02\pm 0.05)\mu_N}{(0.613\pm 0.004)\mu_N}=3.3\pm 0.1,
\label{eq.44}
\end{eqnarray}
 
\noindent so we consistently invoke the SU(3) value to compute
the weak SQL scale Eq. \ref{eq.43}, not Eq. \ref{eq.41}, and use the 
difference between Eqs. \ref{eq.42} and \ref{eq.43}
to give a feel for the error.
 
Returning to $\Omega^-\rightarrow\Xi^0\pi^-$ weak decay, we next
extract the dominant parity-conserving amplitude E from experiment
\cite{pdg,scavis} with cm momentum $p=294$ MeV/c:
 
\begin{eqnarray}
\mid E(\Omega^-\rightarrow\Xi^0\pi^-)\mid=
\frac{m_{\Omega}}{m_{\Omega}+m_{\Xi}}\left[24\pi\Gamma_{\Omega\Xi\pi}
/p^3\right]^{1/2}=(1.33\pm 0.02)\times 10^{-6}~{\rm GeV}^{-1}.
\label{eq.45}
\end{eqnarray}
 
\noindent This observed amplitude in Eq. \ref{eq.45} has remained 
unaltered
for over a decade \cite{uppal,scavis}. It then predicts from pion
PCAC (recall the PCAC structures of Eqs. \ref{eq.1})
 
\begin{eqnarray}
\bigg|\frac{\langle\Xi^0\pi^-\mid H_W\mid\Omega^-\rangle}{
\langle\Xi^-\mid H_W\mid\Omega^-\rangle}\bigg|=
\bigg|\frac{E(\Omega^-\Xi^0\pi^-)}{h_2}\bigg|\approx 
\frac{1}{\sqrt{2}f_{\pi}}~{\rm or}~
\mid h_2\mid\approx 17.5\times 10^{-8}.
\label{eq.46}
\end{eqnarray}
 
\noindent This estimate Eq. \ref{eq.46} is further supported 
using SU(6) Thirring product wave functions \cite{uppal,scavis}. 
Finally, we use Eqs. \ref{eq.43}, \ref{eq.46} to provide the 
third determination of the SQL scale $\beta_W$:
 
\begin{eqnarray}
\mid\beta_W\mid=\mid h_2\mid/3\approx 17.5\times 10^{-8}/3=
(5.8\pm 0.6)\times 10^{-8},
\label{eq.47}
\end{eqnarray}
 
\noindent reasonably near the prior estimates of Eqs. \ref{eq.30},
\ref{eq.39}.
As mentioned above, the error quoted reflects the consistency
indicated by the difference between Eqs. \ref{eq.42} and \ref{eq.43}.
This gives, with Eq. \ref{eq.27},
 
\begin{eqnarray}
\mid\langle\pi^0\mid H_W\mid K_L\rangle\mid =
2\beta_Wm^2_{K_L}f_K/f_{\pi}=
(3.5\pm 0.4)\times 10^{-8}~{\rm GeV}^2,
\label{eq.48}
\end{eqnarray} 
 
\noindent again compatible with the experimental values found in
sect. II.

\vspace{1.0cm}

\noindent {\bf V. DOMINANT LD $K_L\rightarrow\mu\bar{\mu}$ DECAY}
 
\vspace{0.05in}

\begin{figure}[h]
\hspace{3.0cm}
\epsfig{file=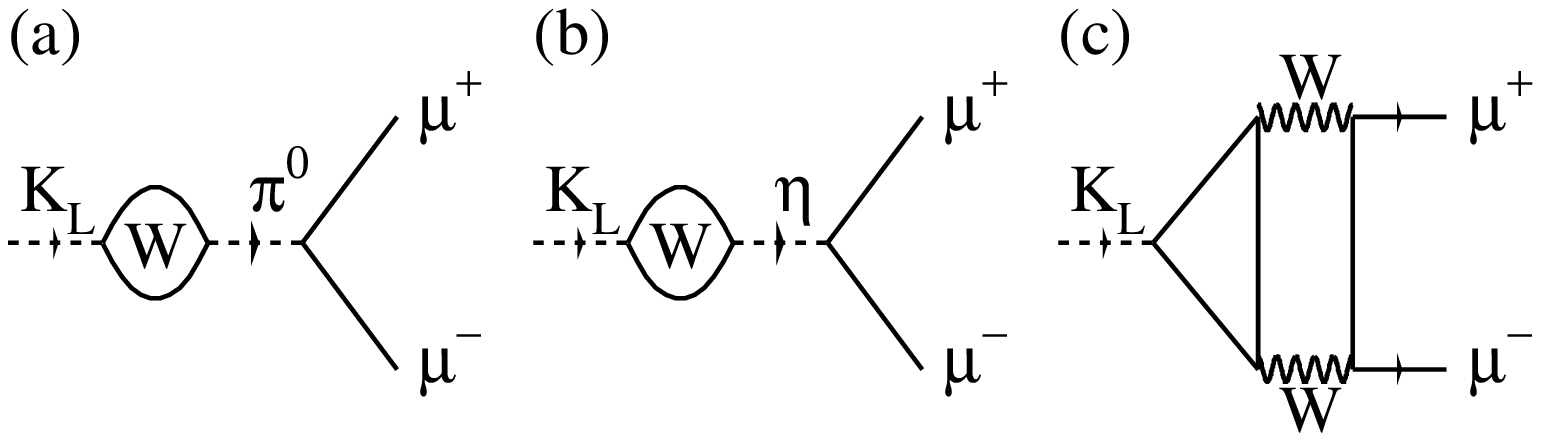,height=6.0cm,width=11.8cm}
\vspace{-2.0cm}
\caption{{\it Weak decay graphs for} $K_L\rightarrow\mu\bar{\mu}$:
(a) LD $\pi^0$ {\it pole}, (b) LD $\eta$ {\it pole},
(c) SD {\it weak box.}\label{fig.8}}
\end{figure}

The dominant first-order weak LD $\pi^0$ and $\eta$ poles are displayed
in Figs. \ref{fig.8}(a), \ref{fig.8}(b) 
while the (smaller) second-order weak short-distance
(SD) graph is shown in Fig. \ref{fig.8}(c). First from 
$\pi^0\rightarrow e\bar{e}$, $\eta\rightarrow\mu\bar{\mu}$, 
$K_L\rightarrow\mu\bar{\mu}$ decay data we extract the various
dimensionless amplitudes defined from an S-matrix element
$S_{fi}=F(P\l\bar{\l})\bar{u}i\gamma_5v\bar{\delta}^4(p_f-p_i)$ 
with lepton spinors normalised covariantly, generating decay rates 
$\Gamma(P\l\bar{\l})=p\mid F(P\l\bar{\l})\mid^2/4\pi$ for cm 
momenta $p=67$ MeV/c, 252 MeV/c, 225 MeV/c for
$\pi^0\rightarrow e\bar{e}$, $\eta\rightarrow\mu\bar{\mu}$,
$K_L\rightarrow\mu\bar{\mu}$ respectively. The PDG tables \cite{pdg} 
then require the central value amplitude magnitudes to be:
 
\begin{eqnarray}
\mid F(\pi^0e\bar{e})\mid=(3.02\pm 0.33)\times 10^{-7},
\nonumber \\
\mid F(\eta\mu\bar{\mu})\mid=(1.85\pm 0.31)\times 10^{-5},
\nonumber \\
\mid F(K_L\mu\bar{\mu})\mid=(2.26\pm 0.05)\times 10^{-12}.
\label{eq.49}
\end{eqnarray}
 
\noindent While the $\pi e\bar{e}$ and $\eta\mu\bar{\mu}$ amplitudes 
in Eq. \ref{eq.49} are of electromagnetic origin \cite{sdrell,sv}, 
it is clear that the much smaller $K_L\mu\bar{\mu}$ amplitude in
Eq. \ref{eq.49} is a weak decay suppressed by $10^{-7}$. 
 
In order to apply the observed amplitudes of Eq. \ref{eq.49} to 
the graph of Fig. \ref{fig.8}(a), we must first scale up the 
$\pi^0e\bar{e}$ amplitude
to the $\pi^0\mu\bar{\mu}$ {\it via} one power of lepton mass (no
covariant normalisation) $m_{\mu}/m_e\approx 206.77$ to
 
\begin{eqnarray}
\mid F(\pi^0\mu\bar{\mu})\mid=206.77\times 3.02\times 10^{-7}=
(6.24\pm 0.68)\times 10^{-5}.
\label{eq.50}
\end{eqnarray}
 
\noindent Then the sum of the graphs Figs. \ref{fig.8}(a),(b) 
predicts the weak amplitude magnitude
 
\begin{eqnarray}
\mid F(K_L\mu\bar{\mu})\mid&=&\bigg|\frac{\langle\pi^0\mid H_W\mid K_L
\rangle F(\pi^0\mu\bar{\mu})}{m^2_K-m^2_{\pi}}+
\frac{\langle\eta\mid H_W\mid K_L
\rangle F(\eta\mu\bar{\mu})}{m^2_K-m^2_{\eta}}\bigg|
\nonumber \\
&\approx& (9.74-7.37)\times 10^{-12}=(2.4\pm 1.6)\times 10^{-12}.
\label{eq.51}
\end{eqnarray}
 
\noindent Here we have taken $\langle\pi^0\mid H_W\mid K_L
\rangle\approx 3.58\times 10^{-8}~{\rm GeV}^2$ as the average
LD weak scale in Sec. II and also divided by $\sqrt{3}$ in 
$\langle\eta\mid H_W\mid K_L\rangle$ using SU(3) symmetry for
$\eta=\eta_8$ since $d_{866}/d_{366}=1/\sqrt{3}$. We find it
significant that the simple estimate in Eq. \ref{eq.51} is so close to
$K_L\mu\bar{\mu}$ data $2.26\times 10^{-12}$ in Eq. \ref{eq.49}. Prior
studies used complex two-loop graphs and $\gamma\gamma$ unitarity
integrals, but still ended up with a result near Eq. \ref{eq.51} anyway
\cite{sdrell,sv}. 
 
Alternatively, we could input the experimental value of the
$K_L\rightarrow\mu\bar{\mu}$ amplitude, 
$(2.26\pm 0.05)\times 10^{-12}$.
If we assume the SU(3) value for $\eta=\eta_8$ of $1/\sqrt{3}$ for 
the ratio $\langle\eta\mid H_W\mid K_L\rangle/
\langle\pi^0\mid H_W\mid K_L\rangle$, we can solve Eq. \ref{eq.51} for
the weak matrix element, giving

\begin{eqnarray}
\mid\langle\pi^0\mid H_W\mid K_L\rangle\mid =
(3.4\pm 2.4)\times 10^{-8}~{\rm GeV}^2.
\label{eq.52}
\end{eqnarray} 

Given the 75\% cancellation between the $\pi^0$ and $\eta$ poles
in Eq. \ref{eq.51}, we must verify that this net LD amplitude is still 
larger than the second-order weak SD amplitude of Fig. \ref{fig.8}(c). 
The latter SD box graph \cite{il} is driven by the heavy t quark at
$m_t\sim 175$ GeV or $X_t=m^2_t/m^2_W\approx 4.8$ with
 
\begin{eqnarray}
G(X_t)=\frac{3}{4}\left[\frac{X_t}{X_t-1}\right]^2{\rm ln}X_t+
\frac{X_t}{4}
+\frac{3}{4}\frac{X_t}{1-X_t}\sim 2,
\label{eq.53}
\end{eqnarray}
 
\noindent then predicting the SD amplitude
 
\begin{eqnarray}
\mid F^{SD}_{K_L\mu\bar{\mu}}\mid\sim s_1c_1s^2_2\times 10^{-9}
G(X_t)\sim 2\times 10^{-13}
\label{eq.54}
\end{eqnarray}
 
\noindent for $s_1c_1\sim 0.22$, $s_2\sim 0.02$. The ratio of
Eq. \ref{eq.51} to Eq. \ref{eq.54} then suggests
 
\begin{eqnarray}
\big|F^{LD}/F^{SD}\big|_{K_L\mu\bar{\mu}}\sim 23.7/2\sim 12,
\label{eq.55}
\end{eqnarray}
 
\noindent so indeed the LD amplitude is more than an order of
magnitude greater than the $K_L\rightarrow \mu\bar{\mu}$ 
SD amplitude.

\vspace{1.0cm}

\noindent {\bf VI. LD $K_S\rightarrow 2\gamma$ WEAK DECAY}

\vspace{0.05in}
 
Finally, we consider the parity-violating (PV) weak decay
$K_S\rightarrow 2\gamma$. Note that the relative errors are
significantly larger than for the parity-conserving (PC) weak
decay $K_L\rightarrow 2\gamma$, i.e. with observed amplitude magnitudes
\cite{pdg,pol} found from $\Gamma(K\gamma\gamma)=m^3_K\mid 
F_{K\gamma\gamma}\mid^2/64\pi$: 
 
\begin{eqnarray}
\mid F_{K_L\gamma\gamma}\mid=
(3.49\pm 0.05)\times 10^{-9}~{\rm GeV}^{-1},~~
\mid F_{K_S\gamma\gamma}\mid=
(5.4\pm 1.0)\times 10^{-9}~{\rm GeV}^{-1},
\label{eq.56}
\end{eqnarray}
 
\noindent with both amplitudes scaled to the Levi-Civita
$\epsilon$ factor.

\begin{figure}[!h!]
\hspace{2cm}
\epsfig{file=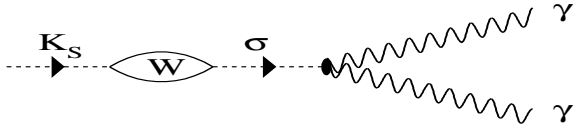,height=7.0cm,width=12.0cm}
\vspace{-1.0cm}
\caption{$\sigma$ {\it pole for} $K_S$ {\it decay.}\label{fig.9}}
\end{figure}

\noindent Given the $\pi^0$ and $\sigma$ pole graphs of Figs. 
\ref{fig.2} and \ref{fig.9}, the corresponding Feynman amplitude 
magnitudes are
 
\begin{eqnarray}
\mid\langle 2\gamma\mid H^{PC}_W\mid K_L\rangle\mid=
\bigg|\frac{\langle\pi^0\mid H^{PC}_W\mid K_L\rangle}
{m^2_K-m^2_{\pi^0}} F_{\pi^02\gamma}\bigg|
\approx 3.49\times 10^{-9}~{\rm GeV}^{-1},
\label{eq.57}
\end{eqnarray}

\begin{eqnarray}
\mid\langle 2\gamma\mid H^{PV}_W\mid K_S\rangle\mid=
\bigg|\frac{\langle\sigma\mid H^{PV}_W\mid K_S\rangle}
{m^2_K-m^2_{\sigma}+im_{\sigma}\Gamma_{\sigma}}\bigg|
\mid F_{\sigma 2\gamma}\mid
\sim 5.4\times 10^{-9}~{\rm GeV}^{-1}.
\label{eq.58}
\end{eqnarray}
 
Although the $\pi^0\rightarrow 2\gamma$ amplitude is actually
measured as $(0.025\pm 0.001)~{\rm GeV}^{-1}$ (in agreement with
the PVV quark graph or the AVV anomaly $\alpha/\pi f_{\pi}=
0.025~{\rm GeV}^{-1}$ with rate $\Gamma_{\pi^0\gamma\gamma}=
m^3_{\pi}\mid F_{\pi^0\gamma\gamma}\mid^2/64\pi$), neither the broad 
$\sigma$ mass-width nor the $\sigma\rightarrow 2\gamma$ decay
rate is accurately known.
              
Taking $\Gamma_{\sigma}\sim m_{\sigma}$ along with the 
(reasonable) chiral value \cite{pol}
 
\begin{eqnarray}
\mid\langle\sigma\mid H^{PV}_W\mid K_S\rangle\mid=
\mid\langle\pi^0\mid H^{PC}_W\mid K_L\rangle\mid\approx
3.58\times 10^{-8}~{\rm GeV}^2
\label{eq.59}
\end{eqnarray}
 
\noindent as found in Sec. II, the estimate \cite{bog}
$\Gamma_{\sigma\gamma\gamma}=(3.8\pm 1.5){\rm keV}$ along with 
the analogue rate
$\Gamma_{\sigma\gamma\gamma}=m^3_{\sigma}\mid F_{\sigma\gamma\gamma}
\mid^2/64\pi$ and Eq. \ref{eq.58} in turn requires the 
central value of the
$\sigma$ mass and the $\sigma\rightarrow \gamma\gamma$ amplitude
to be ({\it without} making any theoretical assumptions)
 
\begin{eqnarray}
m_{\sigma}\approx 614~{\rm MeV},~~\mid F_{\sigma\gamma\gamma}\mid
\approx 2.3\alpha/\pi f_{\pi}\approx 0.057 ~{\rm GeV}^{-1}.
\label{eq.60}
\end{eqnarray}

>From a phenomenological viewpoint, a 614 MeV $\sigma$ mass is
near the central mass now listed in the PDG tables \cite{pdg},
$m_{\sigma}=400-1200~{\rm MeV}$, also close to the E791 collaboration
\cite{791} weak decay analysis, $m_{\sigma}\sim 500 ~{\rm MeV}$.
Moreover, the $\mid F_{\sigma\gamma\gamma}\mid\sim 2.3\alpha/
\pi f_{\pi}$ amplitude value is near the constituent-quark-model 
value of
$(\frac{4}{9}+\frac{1}{9})N_c\alpha/\pi 
f_{\pi}=(5/3)~~\alpha/\pi f_{\pi}$
enhanced by 30\% due to meson $\pi^+$ and $K^+$ loops to
$2.2\alpha/\pi f_{\pi}$ \cite{pol}. It is satisfying that Eqs. 
\ref{eq.60} (resulting from Eqs. \ref{eq.56} - \ref{eq.59}) 
are so close to these experimental
and theoretical values.
 
We end this $\sigma$-dominated $K_S\rightarrow 2\gamma$ section 
by including the $\sigma$-dominated $K_L\rightarrow \pi^02\gamma$
rate. Folding in the latter 3-body phase space integral from
Ref. \cite{ksnc} of $1.7\times 10^{-4}~{\rm GeV}^4$, the PCAC
rate ratio is predicted to be
 
\begin{eqnarray}
\frac{\Gamma(K_L\rightarrow\pi^02\gamma)}{\Gamma(K_S
\rightarrow 2\gamma)}\bigg|_{PCAC}=
\frac{(m_{\sigma}\Gamma_{\sigma})^2(1.7\times 10^{-4}~
{\rm GeV}^4)}{m^6_K(4\pi f_{\pi})^2}=
11.6\times 10^{-4}
\label{eq.61}
\end{eqnarray}
 
\noindent given $f_{\pi}=93 ~{\rm MeV}$, for 
$m_{\sigma}=\Gamma_{\sigma}=614 ~{\rm MeV}$ from
Eq. \ref{eq.60}, close to the measured PDG rate ratio \cite{pdg,pol}

\begin{eqnarray}
\frac{\Gamma(K_L\rightarrow\pi^02\gamma)}{\Gamma(K_S
\rightarrow 2\gamma)}\bigg|_{PDG}= 
\frac{(\hbar/\tau_L)(1.68\pm 0.10)\times 10^{-6}}
{(\hbar/\tau_S)(2.4\pm 0.9)\times 10^{-6}}=
(12.1\pm 4.6)\times 10^{-4},
\label{eq.62}
\end{eqnarray}
 
\noindent for the observed 
lifetimes $\tau_L=(5.17\pm 0.04)\times 10^{-8}{\rm sec}$,
$\tau_S=(0.8935\pm 0.0008)\times 10^{-10}{\rm sec}$.
Note that the weak scale $\langle\sigma\mid H^{PV}_W\mid K_S\rangle$
divides out of the PCAC ratio Eq. \ref{eq.61}. Yet the nearness of
Eqs. \ref{eq.61}, \ref{eq.62} gives further PCAC support for the 
$K_S\rightarrow 2\pi$ and $K_L\rightarrow 3\pi$ PCAC
$K\rightarrow\pi$ transition in Sect. II.

\vspace{1.0cm}
               
\centerline{\bf VII. CONCLUSION}
 
\vspace{0.05in}
 
In Sec.II we showed that the measured decay rates for 
$K_S\rightarrow 2\pi$, $K\rightarrow 3\pi$, 
$K_L\rightarrow 2\gamma$, $K^+\rightarrow\pi^+e\bar{e}$, 
$K^+\rightarrow\pi^+\mu\bar{\mu}$
lead to the average long distance (LD) weak scale 
$\mid\langle\pi\mid H_W\mid K\rangle\mid$ and also derive a
theoretical estimate for this quantity. Then in
Sec.III we computed the single quark line (SQL) s$\rightarrow$d 
dimensionless weak scale $\mid\beta_W\mid$ 
from both first-order and second-order weak transitions.
In Sec.IV we verifed the above weak 
scales by reviewing the observed spin 3/2 
$\Omega^-\rightarrow\Xi^0\pi^-$ weak decays 
and in Sec.V we computed LD and SD $K_L\rightarrow\mu\bar{\mu}$ weak 
decay amplitudes. Finally in Sec VI we studied LD  
$K_S\rightarrow 2\gamma$ weak decay and its PCAC extension
to $K_L\rightarrow\pi^02\gamma$.
 
All of these estimates give strikingly similar values for the weak
$K\rightarrow\pi$ matrix element (without any arbitrary parameters).
Table \ref{tab.1} summarises the results. The average of the seven
experimental values (derived from eleven measured rates)
in Table \ref{tab.1} is 
 
\begin{eqnarray}
\mid\langle\pi^+\mid H_W\mid K^+\rangle\mid=
\mid\langle\pi^0\mid H_W\mid K_L\rangle\mid=
(3.59\pm 0.05)\times 10^{-8}~{\rm GeV}^2,
\label{eq.63}
\end{eqnarray}
 
\noindent where the error quoted is the external error.

\begin{center}
\begin{table}
\center{
\begin{tabular}{|l|c|c|}   
\hline
~~~~~~~~~~~~Source & & Value \\
& & (in units of $10^{-8}~{\rm GeV}^2$)\\ \hline
$K_S\rightarrow 2\pi$ & Expt &                         $3.637\pm 0.009$ \\
$K\rightarrow 3\pi$ & Expt &                           $3.449\pm 0.015$ \\
$K_L\rightarrow 2\gamma$ & Expt &                      $3.56\pm 0.04$ \\
$K^+\rightarrow \pi^+e^+e^-$ & Expt &                  $3.62\pm 0.36$ \\
$K^+\rightarrow \pi^+\mu^+\mu^-$ & Expt &              $4.0\pm 0.6$ \\
$K_L\rightarrow\mu^+\mu^-$ & Expt &                    $3.4\pm 2.4$ \\ 
$\Omega^-\rightarrow\Xi\pi$&Expt&                      $3.5\pm 0.4$ \\
Second-order weak SQL $\Delta m_{LS}$ & Theory &       $3.578\pm 0.004$ \\
W-mediated self-energy graphs & Theory &               $3.47\pm 0.05$ \\ \hline
\end{tabular}
}
\caption{\label{tab.1} Values derived for the $K\rightarrow\pi$
matrix element magnitude.}
\end{table}
\end{center}
 
The errors quoted in Table \ref{tab.1} result mostly from propagation 
of the experimental errors on the input data and,
apart from a few cases, do not include any contribution from
the uncertainty in the theoretical methods. An estimate of the
reliability of the procedures used here
is provided by the external error on the seven experimental
values in Table \ref{tab.1}. The fact that this external error,
quoted in Eq. \ref{eq.63}, is just 1.4\% supports the overall
consistency of the methods used and therefore the reliabilty
of our numerical value for the $K\rightarrow\pi$ weak matrix element.
It is also noteworthy that the experimental result in Eq.
\ref{eq.63} is in excellent agreement with the mean of the
theoretical estimates in Table \ref{tab.1}, which is 
$(3.577\pm 0.004)\times 10^{-8}~{\rm GeV}^2$.

\vspace{0.3in}
 
\centerline{\bf ACKNOWLEDGEMENTS}
 
\vspace{0.05in}
 
We are grateful to G. Eilam, V. Elias and B. Bassalleck for 
many discussions. We acknowledge support from the US DOE.


\begin{thebibliography}{99}
\bibitem{wpole}
The $W$ pole term was added rather than subtracted from 
$\langle\pi^+\mid H_W\mid K^+\rangle$ in Ref. \cite{ks}.
\end{thebibliography}
\end{document}